\documentclass[aps,twocolumn]{revtex4}  
\usepackage{graphicx}  
\usepackage{dcolumn}   
\usepackage{amsmath,amssymb}   
\DeclareMathOperator{\erf}{erf}

\begin{document}

\title{Bubble Chambers for Experiments in Nuclear Astrophysics }  
\author{B. DiGiovine}
\address{Physics Division, Argonne National Laboratory, Argonne, IL 60439, USA}
\author{D. Henderson}
\address{Physics Division, Argonne National Laboratory, Argonne, IL 60439, USA}
\author{R. J. Holt}
\address{Physics Division, Argonne National Laboratory, Argonne, IL 60439, USA}
\author{R. Raut} 
\address{UGC-DAE Consortium for Scientific Research, Kolkata Centre LB-8 Sector-III Bidhannagar, Kolkata 700098, India.}
\author{K. E. Rehm} 
\address{Physics Division, Argonne National Laboratory, Argonne, IL 60439, USA}
\author{A. Robinson}
\address{Department of Physics, University of Chicago, Chicago, IL 60637, USA}
\author{A. Sonnenschein} 
\address{Fermi National Accelerator Laboratory, Batavia, IL 60510, USA}
\author{G. Rusev} 
\address{Chemistry Division, Los Alamos National Laboratory, Los Alamos, New Mexico 87545, USA}
\author{A. P. Tonchev} 
\address{Physics Division, Lawrence Livermore National Laboratory, Livermore, California 94550, USA}
\author{C. Ugalde\footnote{email address: cugalde@anl.gov}}
\address{Department of Physics, University of Illinois at Chicago, Chicago, IL 60607, USA}

\begin{abstract}
A bubble chamber has been developed to be used as an active target system 
for low energy nuclear astrophysics experiments. Adopting ideas from dark 
matter detection with superheated liquids, a detector system compatible 
with $\gamma$-ray beams has been developed. This detector alleviates some 
of the limitations encountered in standard measurements of the minute cross 
sections of interest to stellar environments. While the astrophysically 
relevant nuclear reaction processes at hydrostatic burning temperatures
are dominated by radiative captures, in this experimental scheme we measure 
the time-reversed processes. Such photodisintegrations allow us to compute 
the radiative capture cross sections when transitions to excited states of 
the reaction products are negligible. Due to the transformation of phase 
space, the photodisintegration cross sections are up to two orders of 
magnitude higher. The main advantage of the new target-detector system 
is a density several orders of magnitude higher than conventional gas targets. 
Also, the detector is virtually insensitive to the $\gamma$-ray beam 
itself, thus allowing us to detect only the products of the nuclear 
reaction of interest. The development and the operation as well as 
the advantages and disadvantages of the bubble chamber are discussed.

\end{abstract}

\maketitle

\section{Introduction}

All elements of the periodic table with Z$>$3 have been produced through nuclear reactions in the interior of stars via quiescent or explosive burning processes.  The cross sections of these reactions however, are very small (typically pb-fb) so that they are difficult to measure even at the high temperatures and energies occurring in stellar explosions. In stellar environments this is compensated by the large masses and long time scales involved in stellar evolution. The thin targets which must be used in terrestrial reaction studies (typically $\mu$g-mg/cm$^2$) result in very small luminosities and count rates, with the result that up to now only very few astrophysical reaction cross sections have been measured at stellar temperatures. Most of the other reactions are studied at higher energies and the cross sections are then extrapolated towards the astrophysical energies of interest. In this paper we describe a new detector system working with active liquid targets, which for radiative capture reactions gives an increase in luminosity by several orders of magnitude. Together with existing and planned new accelerators this may enable measurements of many reactions of astrophysical interest under stellar conditions.

\section{Astrophysical background}

The radiative capture of hydrogen or helium (i.e. (p,$\gamma$) or ($\alpha,\gamma$) reactions) on light nuclei such as carbon, nitrogen and oxygen are some of the most important processes in stellar nucleosynthesis. These capture reactions have been studied for many years by bombarding targets of carbon, nitrogen or oxygen with intense proton and $\alpha$-particle beams or (in so-called inverse kinematics) by bombarding hydrogen and helium gas targets with heavier particles at energies of a few hundreds of keV/u and detecting the reaction products with suitable detectors.  The thin targets  ($\sim$ 10$\mu$g/cm$^2$) which are required at these energies together with the beam intensities available at present particle accelerators result in low luminosities with count rates that reach typically 1 count/day for cross sections in the pb region. Small improvements of these yields are still possible but quite long running times will nonetheless be required. If the reaction products from e.g. an ($\alpha,\gamma$) reaction are stable, a considerable improvement of the luminosity can be achieved by studying the inverse ($\gamma,\alpha$) process. One improvement in the expected count rate comes from the reciprocity theorem for nuclear reactions which relates the ($\gamma,\alpha$) process to its inverse ($\alpha,\gamma$) reaction \cite{Blatt}. The two cross sections are related via

\begin{equation}
\frac{\sigma(\gamma,\alpha)}{\sigma(\alpha,\gamma)}
=\frac{\omega_{\alpha,\gamma}k^2_{\alpha,\gamma}}{\omega_{\gamma,\alpha}k^2_{\gamma,\alpha}},
\label{eqn:detailedbalance}
\end{equation}

where k$_{\alpha,\gamma}$ and k$_{\gamma,\alpha}$ are the wave numbers for capture and photodisintegration channels, respectively, and $\omega_{\alpha,\gamma}$ and $\omega_{\gamma,\alpha}$ are the associated spin multiplicity factors. In the energy range discussed in this paper, the ratio can provide a gain of approximately two orders of magnitude in cross section. Another gain in luminosity is obtained from the choice of the target. At the corresponding $\gamma$-ray energies for ($\gamma$,$\alpha$) reactions of 5-10 MeV the large range of the incident $\gamma$-rays allows us to use targets with thicknesses of $\sim$1-10 g/cm$^2$, which corresponds to a factor of 10$^{5-6}$ improvement in luminosity. Disadvantages of this method include the limitation of present tunable $\gamma$-ray sources to about 10$^8$ $\gamma$/s and the need for a detector that is insensitive to the incident $\gamma$-rays. The latter has been achieved by the use of superheated liquids in a bubble chamber to detect and measure the reaction products from the ($\gamma,\alpha$) reaction since these detectors have a high insensitivity to $\gamma$-rays. This has been tested to a level of less than 2$\times$10$^{-10}$\cite{Szydagis} . 

One of the most important capture reactions in nuclear astrophysics is the $^{12}$C($\alpha,\gamma$)$^{16}$O reaction, sometimes called the ``holy grail of nuclear astrophysics'', which can be studied with this technique by using a superheated liquid active target bubble chamber operating with an oxygen-containing liquid. While a study of this reaction is planned for the future, we describe in this paper a study of the capture reaction $^{15}$N($\alpha,\gamma$)$^{19}$F via the photodissociation reaction $^{19}$F($\gamma,\alpha$)$^{15}$N using fluorine containing liquids. This reaction is of interest to nuclear astrophysics since fluorine is the least abundant element in the mass range between 11 and 32 as shown by its solar abundance. This suggests that either it is very hard to synthesize or extremely fragile in stellar environments. Various scenarios for the nucleosynthesis of fluorine have been proposed. One includes the neutrino dissociation of $^{20}$Ne in core collapse supernovae \cite{Woosley:1988}. Others suggest that $^{19}$F could be produced both during the thermal pulse phase in the intershell region of Asymptotic Giant Branch (AGB) stars. Another possibility includes hydrostatic burning in the helium shell of Wolf-Rayet stars. In both cases, the nuclear reaction sequence is the same, with the exception that in the AGB star case, the required neutron flux is induced by $^{13}$C nucleosynthesis produced by the mixing of hydrogen from the envelope into the intershell region and captured by the increasingly abundant $^{12}$C nucleus. For the latter part cross section measurements of the $^{15}$N($\alpha,\gamma$)$^{19}$F reactions at astrophysical energies are needed. In this contribution we describe such a measurement using this newly developed detector using the inverse photodissociation process.  

\section{The Bubble Chamber for use in nuclear astrophysics experiments}

The bubble chamber makes use of the instability of superheated liquids against bubble formation for the detection of charged particles from a nuclear reaction. Detectors of this type have been exploited in high-energy physics experiments for more than 50 years when D. A. Glaser suggested their use to visualize the tracks of high-energy charged particles \cite{Glaser}. Later this technique also found applications in low-energy physics for neutron detection \cite{Jordan}. Larger variants of these detectors are currently employed in dark matter searches \cite{Zacek, Girard, 60kg}. 

The most important difference between ``standard'' bubble chambers used in high-energy physics and the new class of bubble chambers is in their mode of operation. The bubble chambers of high-energy physics were pulsed with the beam bunches, which arrived at well defined time intervals. As a result the bubble chamber would only spend a fraction of a second in the superheated state.  In our application the bubble chamber must stay continuously active until a nuclear reaction occurs in the superheated liquid. With the small cross sections of interest, the active time can be minutes or hours, and so special care must be taken to prevent spurious boiling which is not caused by the nuclear reaction of interest. Another important difference between the device 
and experiments proposed here and those of high energy physics is that no tracks are left by 
the particle inducing nucleation. For the experiments relevant to nuclear astrophysics,
the energies of the reaction products are so small that they are stopped in the liquid after 
a few microns. This means that no direct kinematic information can be obtained 
with this detector. Borrowing heavily from successful designs used in the search for dark matter \cite{Bolte,Zacek,60kg}, we have designed and tested a superheated active target system which will be described in the following sections.  

\subsection{Thermodynamics of bubble detectors}

The study of nucleation in a superheated system has a long history, and is still being investigated today \cite{Harper,Gunther,Uline}.
The theory of bubble formation in a superheated liquid (the so-called ``thermal spike'' model) has been discussed in detail by 
Seitz \cite{Seitz} and will not be repeated here. In the following we provide only a few of the equations, which are necessary for understanding the operating principle.  Figure 1 shows the phase diagram for C$_4$F$_{10}$. 

\begin{figure}
\centering
\includegraphics[width=8.7cm]{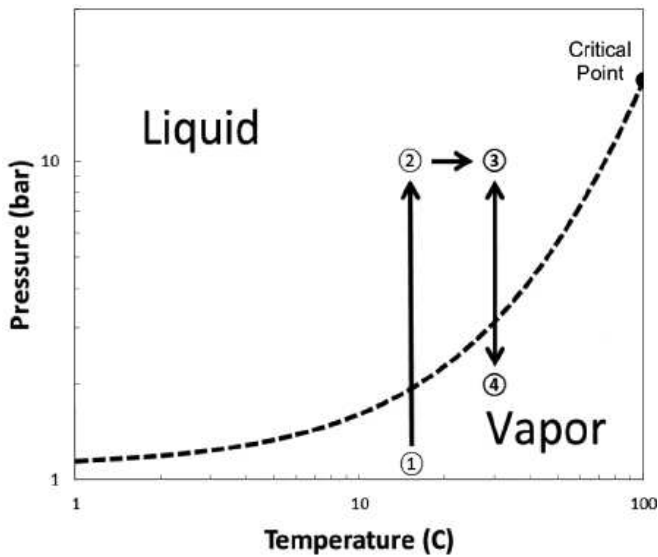}
\caption{Phase diagram of C$_4$F$_{10}$. The solid line represents the path used to bring the fluid into its superheated state}
\label{fig:C4F10PhaseR1}
\end{figure}

The thick solid lines in figure 1 indicate the path that is being used to generate a superheated liquid. Starting at a point where the material is gaseous under ambient pressure and temperature (1), the pressure is first increased, liquifying the gas (2), followed by an increase in temperature (3). Decreasing the pressure below the liquid-gas phase boundary curve brings the fluid into its superheated state (4). For C$_4$F$_{10}$ the operating temperatures are near 30$^\circ$C and the superheat pressures are typically 1-4 bar. If left undisturbed, the fluid will remain liquid in this metastable state. If it experiences a disturbance, which deposits enough energy into the liquid within a certain distance, a critical barrier can be overcome and a ``proto-bubble'' forms. If the ``proto-bubble'' exceeds a critical radius R$_c$ it becomes unstable and the phase transition continues from superheated liquid into the vapor state creating a macroscopic bubble. The critical radius is given by energy conservation arguments as

\begin{equation}
R_c=\frac{2\times\sigma}{(p_v-p_e)},
\end{equation}

where $\sigma$ is the (temperature dependent) surface tension of the liquid, and $\Delta$p= p$_v$-p$_e$ the degree of superheat, i.e. the difference between the vapor pressure p$_v$ at the operating temperature, and the fluid pressure p$_e$. Since the surface tension decreases with temperature, the critical radius R$_c$ decreases towards higher temperatures. Typical values of R$_c$ for the active fluid C$_4$F$_{10}$, discussed in this paper, are between tens and hundreds of nm. 

The energy E$_c$ needed to form a bubble with a critical radius R$_c$, is given \cite{Harper} by the equation

\begin{equation} 
\begin{split} 
E_c=& -\frac{4}{3}\pi R_c^3\Delta p+ \frac{4}{3}\pi R_c^3\rho_v H_{lv}\\&+4\pi R_c^2\left(\sigma-T\frac{d\sigma}{dT}\right) + W_{irr}.
\end{split}
\end{equation}

In equation (3), $\rho_v$ is the density of the material in its gas phase, H$_{lv}$ is the latent heat of evaporation of the superheated liquid and W$_{irr}$ is the energy that goes into irreversible processes such as sound. The first terms in equation (3) correspond, respectively, to the reversible work during the expansion, the energy needed to evaporate the liquid and the energy needed to generate the bubble surface of radius R$_c$. The irreversible contributions expressed by the last term W$_{irr}$ are small and are usually neglected in the calculations. 
\begin{figure}[tbp] 
  \centering
  \includegraphics[width=9.5cm,keepaspectratio]{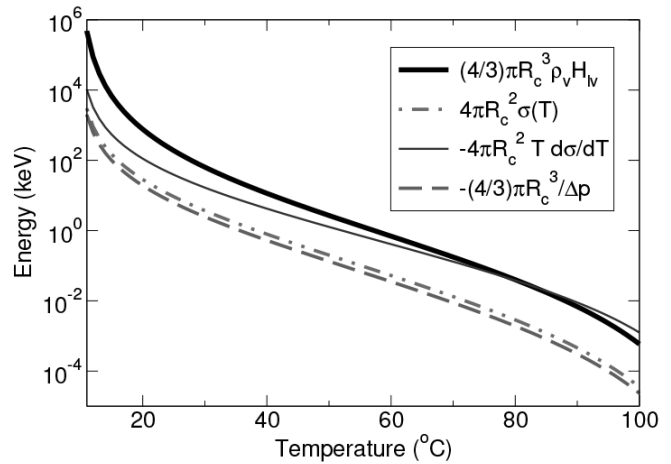}
  \caption{The four terms contributing to the critical energy necessary to form a bubble of critical radius at an operating pressure of 1.82 bar for the 
  liquid C$_4$F$_{10}$.}
  \label{fig:energyTerms_C4F10}
\end{figure}
The various terms contributing to E$_c$ for C$_4$F$_{10}$ are compared in Fig. \ref{fig:energyTerms_C4F10}. The values for the thermodynamic parameters were obtained from the NIST code \cite{REFPROP}. In these calculations, it was assumed that the pressure was 1.82 bar,
which corresponds to the operating value used to superheat the liquid.  As can be seen from Fig. \ref{fig:energyTerms_C4F10}, the dominant contribution to E$_c$ comes from the energy needed to evaporate the liquid, followed by the term originating from the temperature dependence of the surface tension. The contributions from the work needed to generate the surface of a bubble or from the reversible work during the expansion phase are a factor of $\sim$10 smaller.

The ``thermal spike'' for creating a bubble can come from two sources: (1) In the photo-dissociation reaction $^{19}$F($\gamma,\alpha$)$^{15}$N two charged particles, $^{15}$N and an $\alpha$-particle are produced. Depending on the energy of the incoming $\gamma$-ray (5-6 MeV in the experiment described later) the energies are in the range of  200-500 keV for $^{15}$N and between 800-1800 keV for the $\alpha$ particle, respectively. (2) In addition there are background reactions induced by neutrons from cosmic rays. While the liquid in the bubble chamber itself is insensitive to neutrons (neutrons themselves are unable to trigger nucleation), it can be triggered by the $^{12}$C or the $^{19}$F recoils produced in an elastic collision. For a 1 MeV neutron, these recoil energies are in the range of $\sim$300 keV($^{12}$C) or 200 keV ($^{19}$F), respectively, scaling with the energy of the incoming neutron.

In order to generate the thermal spike, these nuclei need to 
deposit this energy over a suitably short distance. This distance 
is somewhat debated in the literature \cite{Archambault} but 
typical values range from 2-6$\times R_c$.
\begin{figure}[tbp] 
  \centering
  \includegraphics[width=9.5cm]{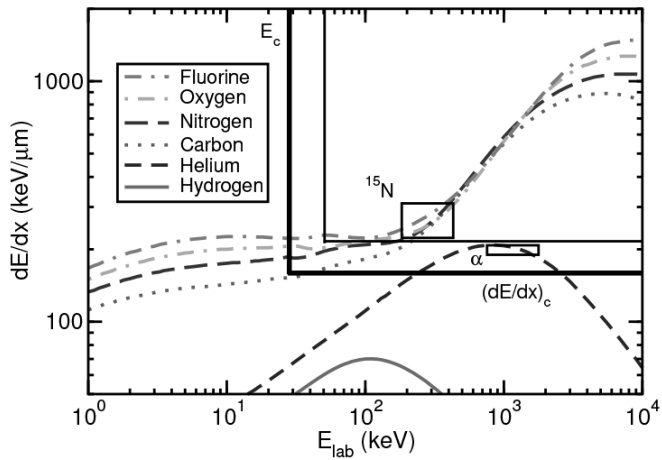}
  \caption{Detection thresholds for the bubble chamber at two different superheats (see text).  The curves represent stopping powers for various nuclei 
  moving in C$_4$F$_{10}$. The liquid will nucleate (form bubbles) for particles in the region delimited 
  by dE/dx$_c$ and E$_c$, which are a function of the superheat in the liquid. These are dependent on the pressure and temperature of the liquid.
  Typical stopping powers and kinetic energies for the $^{19}$F($\gamma,\alpha$)$^{15}$N experiment described in this paper are shown by the small squares for $\alpha$-particles and $^{15}$N. The superheats used for the experiment must define detection thresholds that include the kinematics of the particles as represented in the small squares. When a photodisintegration event produces recoils within the thresholds, a visible bubble will be produced and detected with the CMOS cameras.}
  \label{fig:stoppingC4F10}
\end{figure}

If we integrate the energy-loss curves of the charged particles 
mentioned above over a typical range of 2-6$\times R_c$ we obtain 
the deposited energy E$_{dep}$ which needs to be larger than 
E$_c$ in order to initiate a macroscopic bubble. These conditions 
allow us to select the sensitivity thresholds for particle detection 
by choosing the proper temperatures and pressures (superheat) in the 
bubble chamber. In Fig. \ref{fig:stoppingC4F10} the stopping power 
curves for several charged particles from hydrogen to $^{19}$F in 
C$_4$F$_{10}$ are shown as a function of the energy. Also included are 
the detection limits in dE/dx and E for operating the bubble chamber 
at T=33 $^\circ$C and P=1.82 bar, respectively (thick solid lines). As 
can be seen at these pressures, the bubble chamber is insensitive to 
protons. The detector can also be made insensitive to $\alpha$-particles 
by choosing a smaller amount of superheat, as shown by the thin 
solid lines, which correspond to operating conditions of 
T=30 $^\circ$C and P=1.82 bar.
\begin{figure}[tbp]
  \centering
  \includegraphics[height=11.cm]{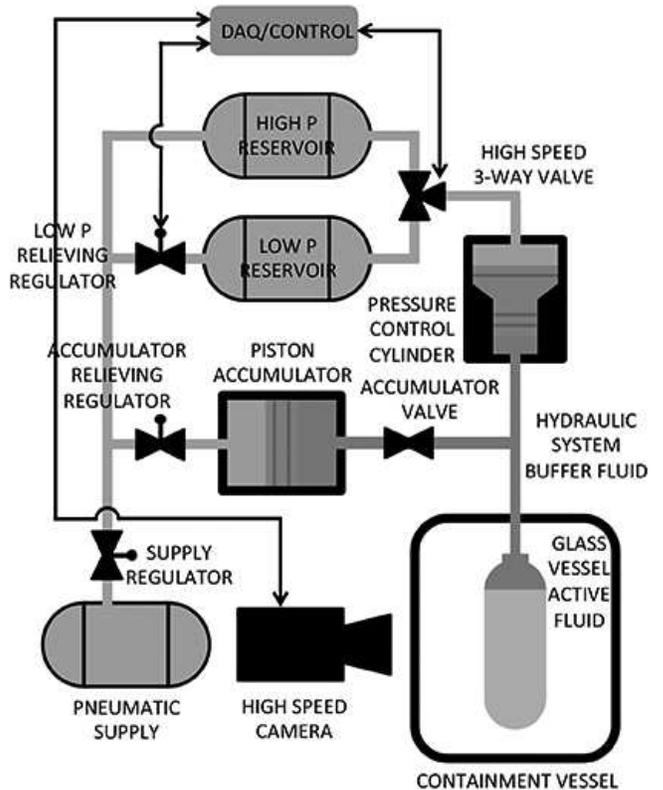}
  \caption{Schematic of the superheated liquid active target system, showing the main sub-systems of the apparatus.}
  \label{fig:Overall}
\end{figure}

\section{Mechanical details}
\begin{figure}[tbp]
  \centering
  \includegraphics[height=6.cm]{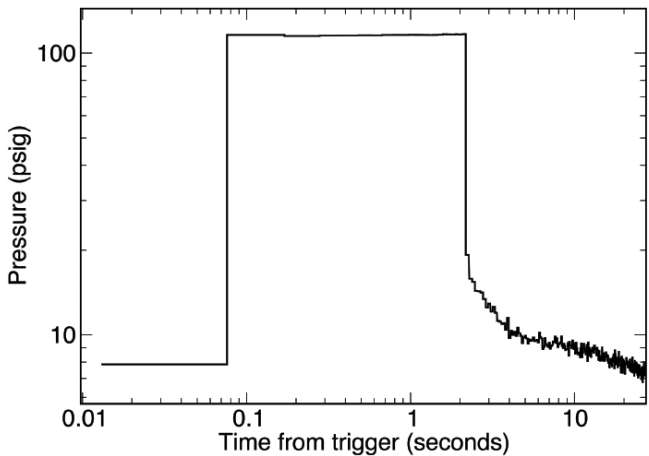}
  \caption{Compression cycle of the liquid after bubble 
  detection. The width of the pressurized state is about 
  two seconds, corresponding to the time it took to liquify 
  the bubble and stabilize the liquid before the superheat 
  was induced again.}
  \label{fig:pressureProfile}
\end{figure}

Figure \ref{fig:Overall} gives an overall schematic of the basic sub-systems of 
the bubble chamber. The containment vessel is filled with the active target 
fluid (bottom) and a so-called buffer fluid (top) fill the volume of the hydraulic system.
The active fluid resides in a high pressure glass vessel allowing 
the less dense buffer fluid to fill the rest of the hydraulic system. 
A piston accumulator compensates 
for volumetric changes during filling and temperature adjustments of 
the active fluid. The hydraulic system pressure is controlled via a 
pneumatically driven pressure control cylinder assembly. 
The pneumatic
system has a high- and low-pressure reservoir, which are connected to the 
pressure control cylinder via a high-speed three way valve. The low- 
pressure reservoir is fed by a precision relieving regulator and sets 
the superheat pressure of the system. The high-pressure reservoir is 
fed directly from the regulated pneumatic supply and sets the compression
pressure, which is typically a factor of 2-3 higher than the vapor pressure 
of the fluid at operating temperature. The glass pressure vessel resides 
within a containment vessel which protects operators and equipment in the 
event of glass failure. High speed CMOS cameras operated by a data
acquisition and control (DAQ$\&$C) system continuously monitor the active 
volume taking pictures every 10 ms, analyzing and capturing events in the 
superheated fluid. The detection of an event is achieved by comparing successive pictures: 
subtracting the individual pixels from one image to the next.
Upon event detection the system is repressurized within
60 ms, which 
liquifies the gas bubble and resets the system, allowing for subsequent 
depressurization back to the superheated state. A compression event is shown in figure \ref{fig:pressureProfile}.

The bubble chamber was operated for several weeks before the experimental campaigns were
started. Cyclic fatigue was not observed as the device was subjected to pressure changes that induced 
the superheat in the liquid.

\subsection{Active target and buffer fluid}

Since in our application the bubble chamber must stay active for extended periods of time, extraneous bubble formation must be kept to a minimum. The active target fluid (C$_4$F$_{10}$) \cite{fluoromed} has to be kept clean and must only contact very smooth surfaces. The fluid is therefore only allowed to come into contact with the glass pressure vessel or with the buffer fluid which provides a smooth interface for the transmission of pressure changes from the hydraulic system. The choice of the buffer fluid depends on several criteria. (1) It must have a lower density so that it floats on top of the active fluid. (2) It must be immiscible with the active fluid in order for a meniscus to form. (3) Solubility between the active fluid and the buffer fluid must be very low, and (4) it should not become superheated in the pressure/temperature range chosen for the experiment. For C$_4$F$_{10}$ as an active fluid, water turns out to be an ideal buffer fluid and was utilized in this experiment.

\subsection{Glass pressure vessel}    
  
The active fluid is contained within a cylindrical glass vessel \cite{chemg} with an overall length of 102 mm, an inner diameter of 30 mm, and an outer diameter of 36 mm. The connection to the hydraulic system is provided by a threaded O-ring sealed bushing. In order to determine the maximum pressure at which this vessel could be operated, several studies were performed. A three-dimensional solid model of the vessel was created using a 3D CAD software package. This model was then used to perform stress analysis simulations, which predicted a failure of the vessel at pressures around 100 bar. Hydrostatic tests were then performed up to 80 bar on several vessels, with no failures observed. Considering simulations and hydrostatic testing, the glass vessel provided our system with a safety factor of about 8-10. Because of the brittle nature of glass and the possibility of stress risers, which can occur in glass due to small scratches or imperfections, the glass vessel was housed within a secondary stainless steel containment vessel.
\begin{figure}[tbp] 
 \centering
  \includegraphics[width=9.cm,keepaspectratio]{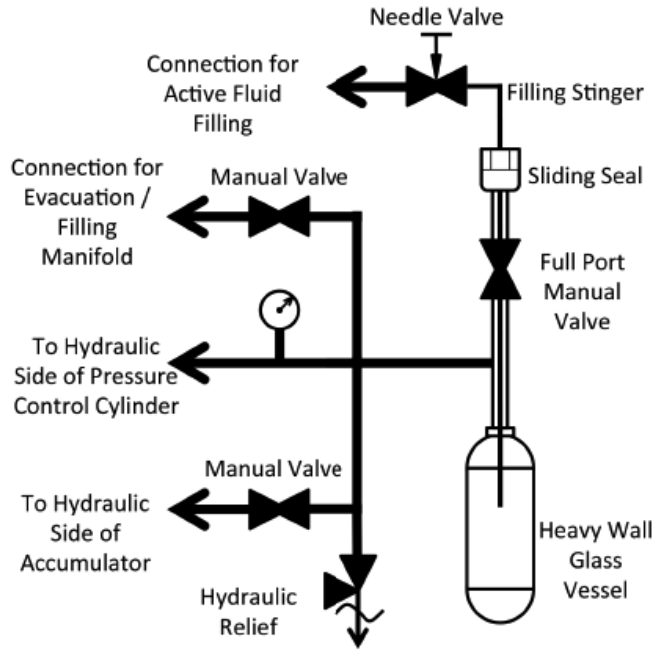}
  \caption{Schematic of the hydraulic system.}
  \label{fig:Hydraulic}
\end{figure}

\subsection{Hydraulic System}

There are several requirements for the hydraulic system. (1) It provides the interface between the superheated liquid and the pneumatic pressure control system, (2) it allows for thermal expansion compensation, and (3) enables the filling of the pressure vessel with the active fluid. A schematic diagram is shown in Fig. \ref{fig:Hydraulic}. The hydraulic system pressure is controlled by utilizing an unbalanced piston and cylinder assembly which is the main interface 
\begin{figure}[tbp] 
  \centering
  \includegraphics[width=8.cm,keepaspectratio]{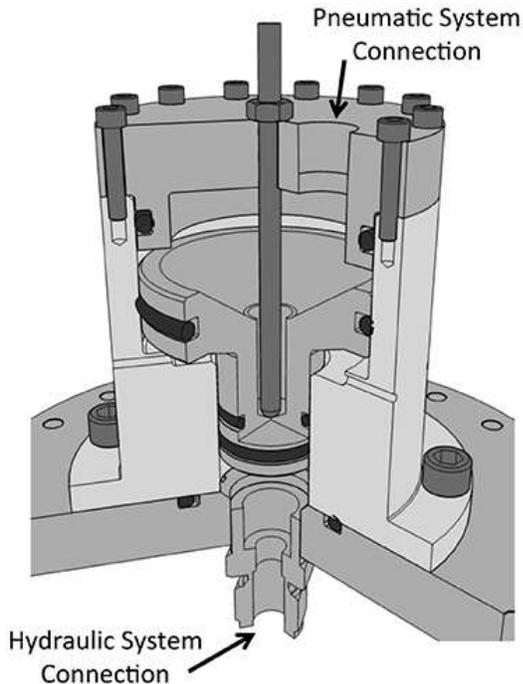}
  \caption{Detail cut-away of the pressure transfer piston and cylinder assembly providing the interface between the pneumatic and hydraulic systems. The lower volume belongs to the hydraulic system, the upper volume is part of the pneumatic system.}
  \label{fig:UnbalPistonR1}
\end{figure}
between the hydraulic and the pneumatic system. The design of this piston assembly is shown in Fig. \ref{fig:UnbalPistonR1}. The hydraulic side of the cylinder has a diameter 
of about 40\% of the pneumatic side, allowing for higher pressures to be reached using standard off-the-shelf pneumatic control valves, regulators and tanks. The travel of the piston in both directions is limited by flanges. This helps to prevent the complete vaporization of the superheated fluid in case of 
an event that proceeds undetected by the video cameras and the computer system.  

An accumulator with an adjustable pneumatic charge is used for thermal expansion compensations and when significant temperature excursions are needed during filling operations. Once the detector is filled and has reached its operating temperature, the accumulator is valved off and put in standby. The hydraulic system pressure is monitored by a fast response ($\ < $1 ms )  pressure transducer \cite{omega}. It is continuously monitored to ensure the correct superheat and recompression pressure. Its value is also recorded for each detected event and event triggering is possible on the pressure signal. This means that as pressure rises due to bubble 
growth, the recompression cycle can be initiated without the need of the optical signal from 
the video cameras.   

A proper filling of the system with the various liquids is critical. Trapped gases in the hydraulic system result in poor pressure response. Small amounts of the active fluid in contact with rough surfaces outside the glass vessel cause pressure control problems as well as rapid boiling during decompression. A typical filling procedure involves the evacuation of the entire hydraulic system to remove all trapped gases. Distilled and degassed water is then allowed into the system through a filter until the entire volume of hydraulic system is filled with water. The accumulator is then used to increase the pressure to slightly below the vapor pressure of the active fluid at room temperature and the hydraulic system is chilled. Once this is complete, gaseous C$_4$F$_{10}$ is slowly allowed into the glass pressure vessel through a filling stinger (see Fig. \ref{fig:Hydraulic}). The gas condenses as it travels down through the stinger and sinks to the bottom of the glass pressure vessel. The active fluid then displaces the buffer fluid in the bottom of the glass vessel, and the change in fluid volume is taken up by the piston accumulator. Visual inspection allows for the determination of the success of the filling process. Once filling is complete, a sliding seal and a ball valve facilitate the removal of the filling stinger. The accumulator is valved out of the hydraulic system and proper pressure response is verified. The accumulator can be valved back in to the system and the temperature can be increased to the operating temperature to prepare the system for operation.
\begin{figure}[tlbp] 
  \centering
  \includegraphics[width=8.5cm,keepaspectratio]{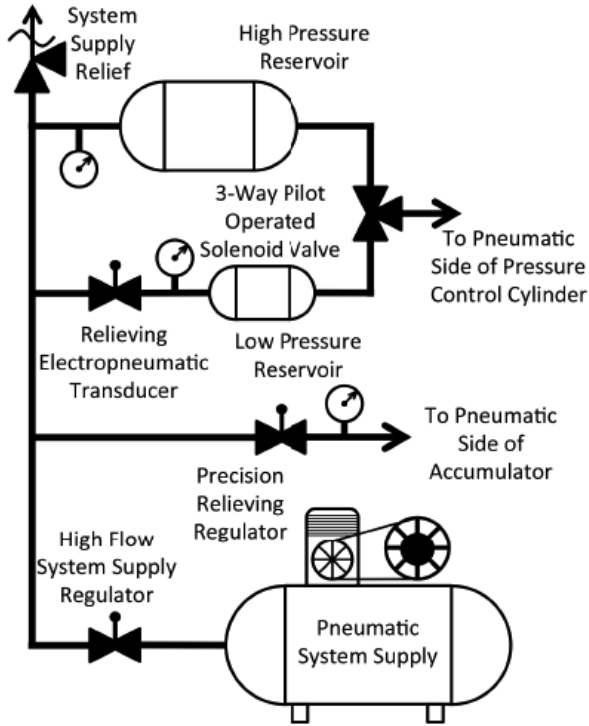}
  \caption{Schematic of the pneumatic system.}
  \label{fig:Pneumatic}
\end{figure}

\subsection{Pneumatic pressure control and temperature monitoring}

The pressure in the hydraulic system is controlled by the pneumatic system as shown in Fig. \ref{fig:Pneumatic}. A fast response (10 ms) high flow rate, three-way pilot-operated pneumatic solenoid valve \cite{MAC} is directly connected to the cylinder head of the pressure control cylinder. This valve is controlled by a signal from the DAQ$\&$C computer, which allows for rapid pressure changes from superheat to compressed conditions. The valve is also connected to two pneumatic tanks which act as reservoirs. The main compression tank is fed directly from a compressed air source and regulated to achieve the proper compression pressure. A connection from this supply is routed to a precision self-relieving regulator, which feeds the superheat pressure tank. This regulator is an electropneumatic transducer \cite{contAir}, which allows for remote control of its output pressure. This enables fine control of superheat pressure without the need to access the device.  

Since the temperature of the system has to be stable to within 1$^\circ$C the interior of the containment vessel is thermally insulated. Several resistance temperature detectors (RTDs) are mounted on the glass vessel to monitor the temperature of the superheated liquid. The signals from the RTDs are fed into temperature controllers which display and control the temperature in the chamber by activating a small heater installed near the bottom of the safety vessel. Thermal gradients across the active volume are smaller than 0.5 $^\circ$C.

\subsection{CMOS cameras, lighting and data acquisition}

\begin{figure}
  \centering
  \includegraphics[width=8.5cm,keepaspectratio]{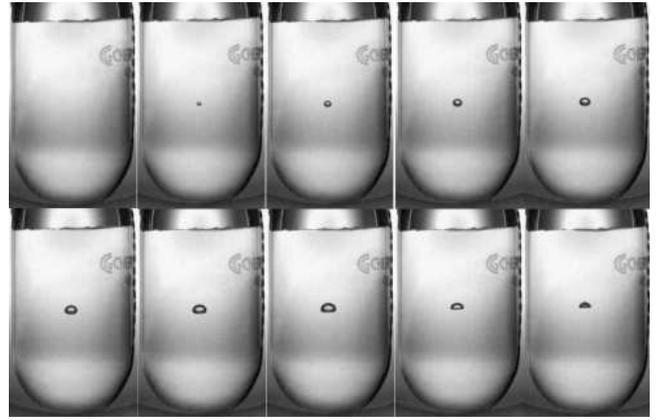}
  \caption{Series of 10 pictures taken by the CMOS camera showing the development of a bubble generated in superheated C$_4$F$_{10}$ via the $^{19}$F($\gamma,\alpha$)$^{15}$N reaction. The time difference between individual pictures is 10 ms.}
  \label{fig:Bubble_Fig_7}
\end{figure}

The containment vessel has two viewing ports located at $\pm$45$^\circ$ with respect to the incident $\gamma$-ray beam. Two 100 Hz high-sensitivity CMOS area scan cameras \cite{basler} are continuously monitoring the active volume of the glass vessel. To increase contrast and sensitivity,  a high intensity diffuse LED backlight is mounted opposite to each camera. One of the cameras acts as master and one as slave. The DAQ$\&$C computer continuously subtracts each new frame provided by the master camera from the previous frame and evaluates the difference. It takes about 5 ms to analyze the frames, which allows us to take full advantage of the speed of the camera (10 ms per frame). Once a difference above a certain threshold has been detected, the system is triggered by sending a signal to the control system to repressurize. A time stamp is assigned to this event and a sequence of 10 consecutive frames from the master camera as well as one frame from the slave camera is written to disk. This provides spatial information about the event within the vessel. Pressure and temperature condition of the bubble chamber at the time of the event are also recorded. As an example Fig. \ref{fig:Bubble_Fig_7} exhibits the
sequence of 10 pictures from the master camera which show the growth and the recompression of one bubble event. The bubble is found to grow for the first 40-50 ms, which is the time needed to (1) recognize a difference in two subsequent frames, (2) send a signal to a fast relay which energizes the pneumatic valve, and (3) pressurize the pneumatic volume of the pressure control cylinder.  This initiates a pressure increase sufficient to quickly liquify the bubble. After an (adjustable) time difference, typically of 1 to 5 s, which is the main component to the dead time (see figure \ref{fig:pressureProfile} for a compression of 2.1 s), the DAQ$\&$C system decompresses the bubble chamber again so that it is ready for the next event.

Given the small growth rate of a bubble during the first 10 ms after nucleation, the bubble chamber is able to resolve the site of bubble formation with a resolution of the order of the resolving power of the video camera. This is well below 1 mm,  i.e. smaller than the size of the $\gamma$-ray beam which is defined by a 10 mm collimator.

\section{Experimental Details}
\subsection{Production of the $\gamma$-ray beam}
Linear or circular polarized $\gamma$-rays with energies ranging from 5 to 10 MeV are produced at the High Intensity $\gamma$-ray Source (HI$\gamma$S) at Duke University \cite{Weller:2009}. A full description of this facility can be found elsewhere and only a few details will be discussed here. The $\gamma$-rays are produced through inverse Compton scattering of a two-bunch electron beam circulating in an electron storage ring with the photons from a high-power free electron laser (FEL) beam. Each electron bunch had different charges: a big one to induce the FEL ($\sim$40 mA) and a small one ($\sim$1 mA) to produce the $\gamma$-ray beam by inverse Compton scattering. This scheme helped reducing the bandwidth of the $\gamma$-ray beam profile. 

The energy of the electrons was typically 400 MeV with a 532 nm wavelength of the laser light. With this technique intensities up to about 10$^8$ $\gamma$/s can be achieved. The $\gamma$-ray beam was collimated to a diameter of 10 mm with a series of oxygen-free Cu collimators. The ($\gamma$,n) thresholds of $^{63,65}$Cu are at 10.85 MeV and 9.90 MeV, respectively, preventing the production of neutrons which, as discussed earlier, can be a source of background in the experiment. For the same reason the entrance window of the safety vessel of the bubble chamber consisted of 3 mm thick Al flanges. At the low $\gamma$-ray energies used in this experiment contributions from ($\gamma$,n) reactions on other isotopes (such as $^{57}$Fe and $^{29}$Si) do not result in any neutron background that trigger the bubble chamber. As will be discussed later, neutrons can be produced through collisions of the electron beam with the residual gas in the storage ring. The flux of the incident $\gamma$-ray beam was controlled by inserting a set of Cu attenuators into the beam path. Typical beam intensities used in the experiment ranged from 2$\times$10$^3$ to 5$\times$10$^6$ $\gamma$/s.  
 
\begin{figure}[tbp]
  \includegraphics[width=10.9cm,keepaspectratio]{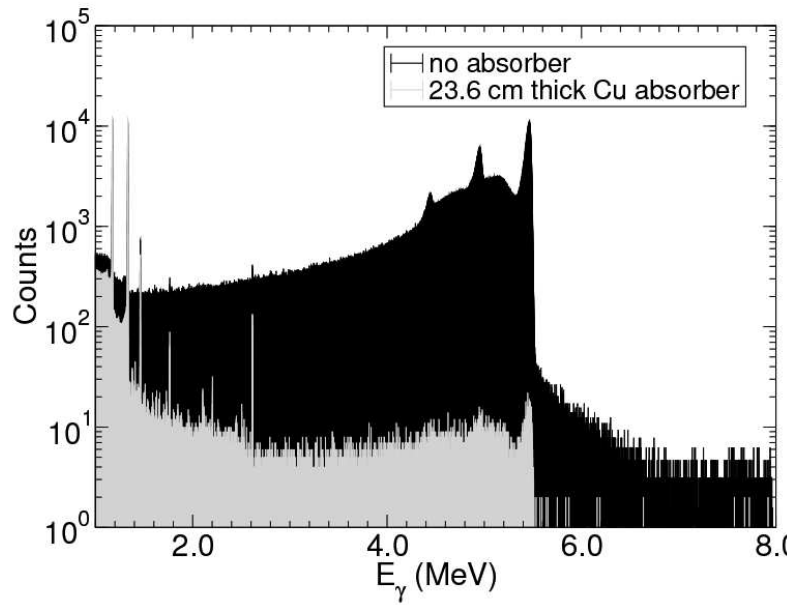}
  \caption{ $\gamma$-ray spectra measured with a 123\% HPGe detector placed downstream of the bubble chamber. Inverse Compton Scattering $\gamma$ rays produced by the HI$\gamma$S facility have a centroid energy of E$_{\gamma}$=5.44 MeV. The spectra show the total count numbers during 39 minutes of beam measurement. The spectrum in black was taken with a 23.6 cm thick Cu absorber between the bubble chamber and the HPGe detector, while the spectrum in gray was acquired with no attenuator in place. Energy calibration for the spectra was performed by placing a $^{60}$Co source next to the HPGe detector. The flux of $\gamma$-rays incident on the target was kept constant at 2.8$\times$10$^3$ $\gamma$/s between both runs. Here, the peaks at 5.44 
MeV correspond to the photopeaks, while those around 4.93 MeV are the first escape peak, and those at 4.42 MeV correspond to the second escape peak.}
  \label{fig:attenuatorSpectrum}
\end{figure}

\subsection{Beam profile}
Two $\gamma$-ray spectra measured with a 123\% efficiency HPGe $\gamma$-ray detector (detection efficiency at 1.33 MeV relative to that of a standard 3-in.-diameter, 3-in.-long NaI(Tl) scintillator) placed $\sim$173 cm downstream after the bubble chamber are shown in Fig. \ref{fig:attenuatorSpectrum}.  The count rate in the detector was kept under 50 kHz by attenuating the flux with 10 cm thick Al slab or a 23.6 cm thick Cu absorber placed between the bubble chamber and the $\gamma$-ray detector. 

The measured spectrum was corrected for the background produced by the $\gamma$-ray beam scattered from the exit window of the 
bubble chamber and the Al or Cu absorber and also corrected for the detector response of the HPGe detector using GEANT \cite{geant} 
simulations. The full energy peak detection efficiency was simulated using the energy spatial correlation of the $\gamma$-rays in 
the beam \cite{Sun} unique for each measurement. The $\gamma$-ray flux was obtained by dividing the background corrected 
spectrum by the efficiency and the live time of the measurement.

An {\it Exponentially Modified Gaussian} (EMG) distribution with the form  
\begin{equation}
\begin{split}
f(E)&=\frac{ac\sqrt{2\pi}}{2d}  
\exp{\left({\frac{b-E}{d}}\right)}
\\
&+{\frac{c^2}{2d^2}}
\left[
{\frac{d}{|d|}}
-\erf{\left(\frac{b-E}{\sqrt{2}c}+\frac{c}{\sqrt{2}d} \right)}
\right]
\end{split}
\end{equation}
was assumed to obtain the measured flux.

\begin{figure}
  \centering
  \includegraphics[width=8.6cm]{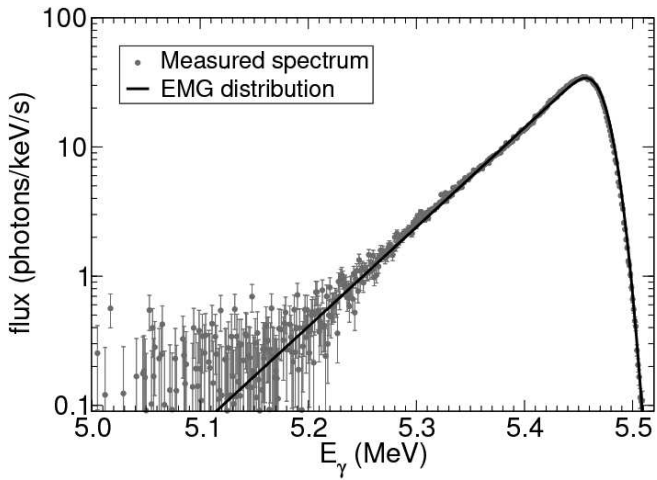}
  \caption{Reconstruction of the incident $\gamma$-ray beam at E$_{\gamma}$=5.44 MeV obtained from a Monte Carlo simulation. The solid line represents the parameterization of the Monte Carlo data. The error bars contain statistical contributions only. An extra 3\% error bar for systematic effects needs to be added to the data.}
  \label{fig:beamSpectrum}
\end{figure}

Here a, b, c, and d are the fitting parameters, and E is the $\gamma$-ray energy.  A comparison of the experimental and the Monte Carlo simulated flux is shown in Fig. \ref{fig:beamSpectrum} with the solid line representing the EMG distribution.

\section{Experimental results}
\begin{figure}
  \centering
  \includegraphics[width=9.1cm,keepaspectratio]{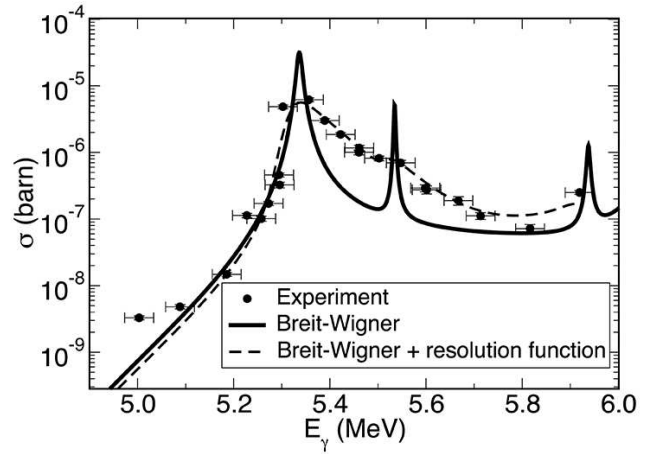}
  \caption{Cross section for the $^{15}$N($\alpha$,$\gamma$)$^{19}$F reaction computed with a Breit-Wigner model using parameters from \cite{Wilmes}. In solid black, the cross section is presented, while in dashed black, the curve folded with the resolution function (see Fig.\ref{fig:beamSpectrum} ) of the beam is shown.}
  \label{fig:BWfolded}
\end{figure}

\begin{figure}
  \centering
  \includegraphics[width=8.2cm,keepaspectratio]{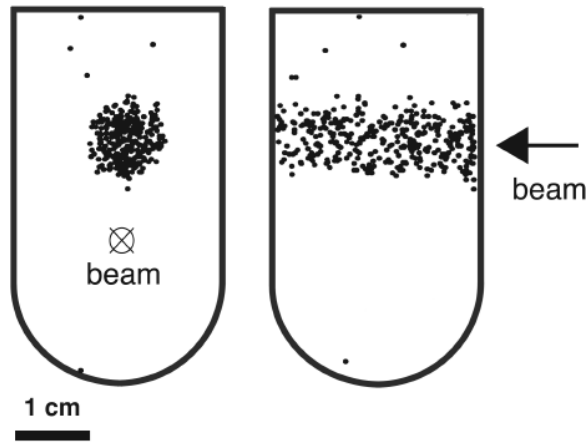}
  \caption{Sites of nucleation for a beam-on-target run. The profile of the 
  $\gamma$-ray beam can be clearly observed. Some background events are also seen. }
  \label{fig:fiducialPlot}
\end{figure}

For a proof-of-principle study of this detector system we have measured the $^{15}$N($\alpha$,$\gamma$)$^{19}$F reaction via photodissociation of $^{19}$F, i.e. through the $^{19}$F($\gamma$,$\alpha$)$^{15}$N reaction \cite{UgaldePLB}. Resonance parameters for the ($\alpha$,$\gamma$) reaction have been measured previously \cite{Wilmes}. Using these values in a Breit-Wigner model we obtain a predicted spectrum shown in Fig. \ref{fig:BWfolded}  plotted as function of the excitation energy in $^{19}$F with cross sections ranging from 100 pb to 30 $\mu$b. In our measurement with the bubble chamber we have studied this excitation energy region with $\gamma$-rays from 5-6 MeV. The beam intensities ranged from 2$\times$10$^3$ to 5$\times$10$^6$ $\gamma$/s depending on the cross section. The typical running time for one energy was about 1 hour. From the information of the two CMOS cameras located  $\pm$45$^{\circ}$ with respect to the incident beam one can calculate the shape of the intersection between the 
beam and the glass vessel.  This is shown in Fig. \ref{fig:fiducialPlot}, where the location of bubbles measured during one run appear looking along and perpendicular to the $\gamma$-ray beam. No corrections for refraction in the walls of the glass vessel have been made as the position 
shifts observed were minimal with and without the superheated liquid. The dimensions of the 
fiducial area are in good agreement with the expectation. The events outside the fiducial volume are present even without a $\gamma$ beam 
and are caused by $^{12}$C and $^{19}$F recoils after elastic scattering of cosmic ray induced neutrons. The neutron-induced background rate in the fiducial volume is about 1$\times$10$^{-3}$ events/s. The number of bubbles, N$_{counts}$, measured at a given energy and corrected for background was then used to calculate the reaction yield, given by

\begin{equation}
Yield=({{N_{counts}}\over{\phi_\gamma}}){{1}\over{\Delta t-t_{dead}*N_{bubble}}}.
  \label{eqn:yield}
\end{equation} 

Here $\phi_{\gamma}$ is the energy integrated flux of $\gamma$-rays hitting the target (see Fig. 10), $\Delta$t the running time, t$_{dead}$ the dead time of the detector discussed earlier (t$_{dead}$=2 s) and  N$_{bubble}$ the number of bubbles observed in the whole active volume of the detector. 

The cross section of the $^{19}$F($\gamma,\alpha$)$^{15}$N reaction was then calculated from the reaction yield in equation 5 by

\begin{equation}
\sigma(\gamma,\alpha)= {\frac{Yield}{L}} {\frac{1}{\nu\rho}} {\frac{N_A}{A}},
\end{equation}

where L is the target thickness (L =3cm), $\nu$ is the number of fluorine atoms per C$_4$F$_{10}$ molecule ($\nu$=10), $\rho$
is the density of the liquid at superheated conditions ($\rho$ = 1.45 g/cm$^3$), N$_A$ is Avogadro's number, and A is the molecular weight of C$_{4}$F$_{10}$ in a.m.u. (A=238).
The cross section of the  $^{19}$F($\gamma,\alpha$)$^{15}$N reaction was then used to calculate $\sigma$($^{15}$N($\alpha,\gamma$)$^{19}$F ) with equation \ref{eqn:detailedbalance}. The results of the cross sections assuming a detection efficiency $\epsilon$=1 are shown as the points in Fig. \ref{fig:bwm} with the appropriate uncertainties in E and $\sigma$. The error bar for the cross section corresponds to the statistical uncertainty, while the error bar for the energy was obtained from the FWHM of the $\gamma$-ray beam as modeled in Fig. \ref{fig:beamSpectrum} and converted to E$_{lab}$ in the $^{15}$N($\alpha,\gamma$)$^{19}$F reference system.

The cross sections range from about 10 $\mu$b at the peak of the resonance to about 3 nb at the lowest energy. At the highest cross section the beam intensity had to be reduced to about 2$\times$10$^3$ $\gamma$/s, confirming the high luminosity of this detector system. The solid line corresponds to the excitation function folded with the resolution function of the beam as described in Fig. \ref{fig:beamSpectrum}. As can be seen an excellent agreement between the experimental data and the theoretical prediction is observed, so the assumption of $\epsilon$=1 for the 
detection efficiency of the reaction products is also confirmed. This is in agreement with the results presented in Ref. \cite{Zacek}.  

\subsection{Beam-induced background}
\begin{figure}
  \centering
  \includegraphics[width=8.9cm,keepaspectratio]{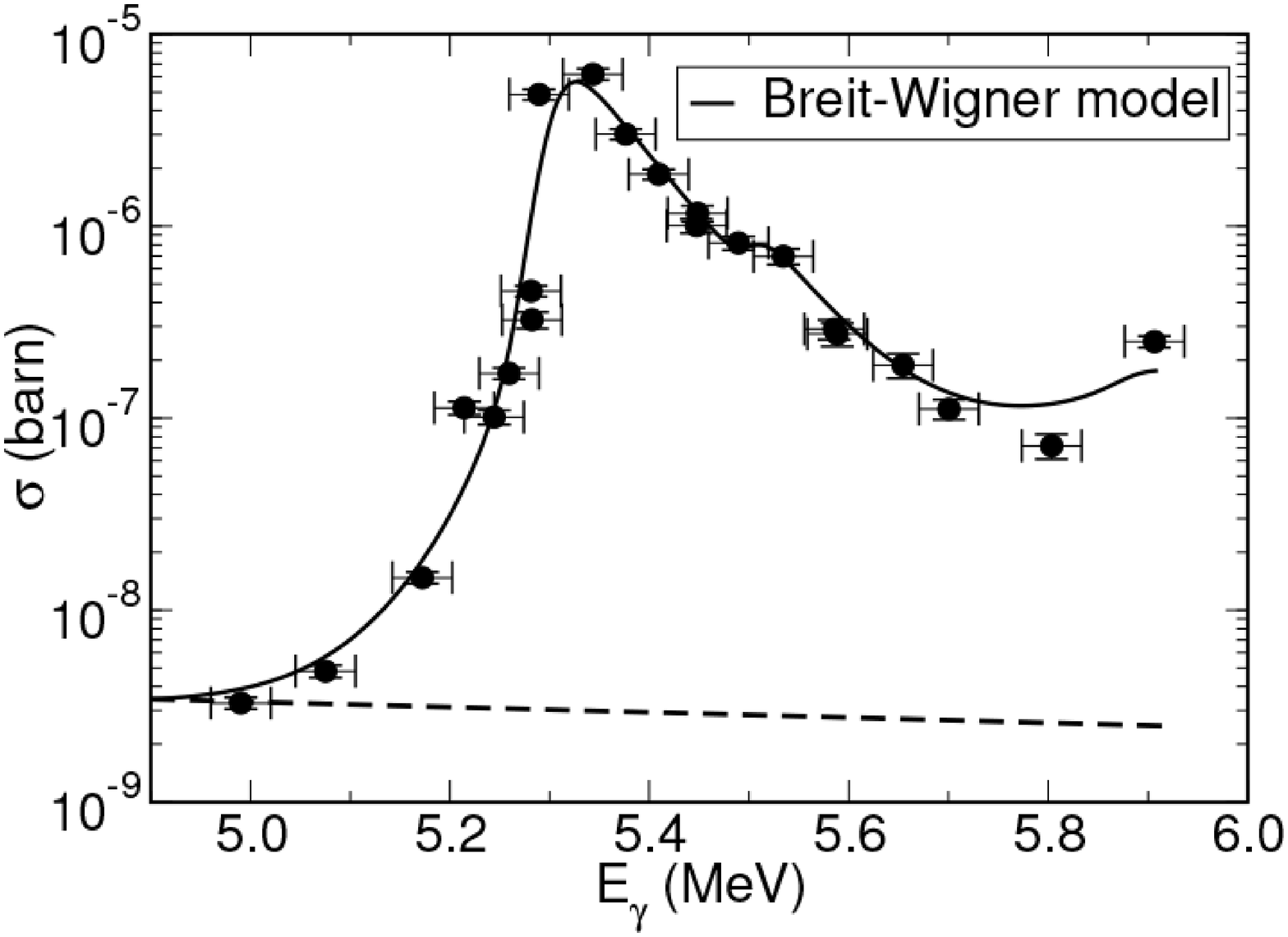}
  \caption{Cross sections calculated with a Breit-Wigner 
  model with resolution function and corrected with a 
  Bremsstrahlung background component. The data points were obtained experimentally 
  and are in good agreement with the model. The dashed 
  curve represents the background while the solid line 
  is the sum of the dashed line of figure \ref{fig:BWfolded} 
  and the Bremsstrahlung background.}
  \label{fig:bwm}
\end{figure}

As can be seen from Fig. \ref{fig:bwm}, the experimental cross sections
at the lowest energies saturate at $\sim$ 3 nb, overestimating the value 
predicted by the Breit-Wigner distribution (see Fig. \ref{fig:BWfolded}). 
The count rate measured at these energies contain a contribution from a 
beam induced background source that has not been accounted for so far. 
These background signals cannot be distinguished from the signal of 
interest from $^{19}$F($\gamma,\alpha$)$^{15}$N, as they appear in the same 
spatial region when the beam is incident on the target. They are caused 
by a Bremsstrahlung radiation component produced by the electrons and 
the residual gas in the beam line. While small in comparison 
with the main $\gamma$-ray beam component, these $\gamma$-rays have energies 
ranging from the electron beam energy (some hundreds of MeV) down to zero. 
Thus even a small amount of high-energy Bremsstrahlung $\gamma$-rays can 
dominate the yield at the lowest energies.

In order to estimate the flux from this background source we assume electrons 
circulating in a storage ring in which the vacuum has a typical value of 
2$\times$10$^{-10}$ Torr. The section of the beam line in the ring where the 
electrons move in the direction of the bubble chamber has a length of 35 m. 
Assuming a beam energy of 400 MeV, an electron beam current of 40 mA, 
Z=10 residual gas \cite{Schreiber},  a 3 cm thick C$_4$F$_{10}$ target with 
($\gamma$,n) cross sections of 15 mb between 15-30 MeV and 0.5 mb 
elsewhere the count rate for this Bremsstrahlung induced background source 
would be about 0.1 counts per second, which is in good agreement with the 
count rate values measured at the lowest cross sections in our experiments.
Converting this background yield into cross sections and plotting it in the 
$^{15}$N($\alpha$,$\gamma$)$^{19}$F plot gives the dashed line in Fig.  
\ref{fig:bwm}. Adding this contribution to the folded Breit-Wigner distribution 
results in an excellent agreement with the experimental data.

There are two possible ways of reducing this background contribution: the
first includes the suppression of the Bremsstrahlung radiation by rearranging the 
accelerator beamlines at the region where the laser-electron collision takes place. 
The electron beam length in the direction of the $\gamma$ rays could be reduced by 
bending their path with a magnet right before the collision point. The second 
possibility includes the separation of neutron-induced events from the heavy ion 
recoils by using the sound produced by the nucleation events. This technique has 
been successfully used in superheated liquid devices built for dark matter 
detection. The different spectral charactersitics of the sound produced by these 
two types of events allow for their separation \cite{Aubin}.

\subsection{Other backgrounds and systematic uncertainties}
Contributions to the count rate that are accounted for as backgrounds and 
are unrelated to the $\gamma$-ray beam or its production process include 
cosmic ray induced neutrons and muons, radioactivity in the vessel from 
the chemical makeup of the glass, and neutrons photodisintegrated 
with $\gamma$-rays produced by the experimental room walls. All these 
backgrounds were measured by operating the bubble chamber 
without an electron beam in the accelerator ring. Cosmic ray
induced events will appear evenly distributed over the whole volume 
of the superheated liquid. They can be reduced by passive shielding 
around the bubble chamber. Radioactivity from the glass only triggers 
nucleation on the superheated liquid in close contact with the vessel walls.
 
These backgrounds can be identified by using the good spatial 
resolution capabilities of the bubble chamber. By knowing the fiducial 
region where the $\gamma$-ray beam irradiates the superheated liquid, 
the majority of these background events can be suppressed. It is only 
those produced in the fiducial region that need to be accounted for by doing 
background runs without beam in the accelerator. In these tests, the average 
time between consecutive nucleation events was of two minutes, while the 
typical count rate of the detector when $\gamma$-ray irradiated it was of 
the order of 0.1 Hz. This results in an average 8.3\% contribution to the 
count rate coming from these sources of background combined. While some 
bubbles were observed next to the glass walls (one or two per hour of 
operation), the contribution to the whole count rate from these was 
insignificant. These nucleation sites could be associated with 
$\alpha$-particle decays from radioactive nuclei in the glass.

An additional background contribution that was studied in this work 
was that originating from photo-induced neutrons produced upstream 
in the accelerator by the interaction of $\gamma$-rays with beam line components.
We measured this contribution by moving 
the bubble chamber to the side out of the $\gamma$ ray beam path 
and determining a count rate. These bubbles were  
distributed evenly in the superheated liquid volume and account 
for 8\% of the count rate measured in runs with the $\gamma$-ray 
beam impinging on the bubble chamber. 

Other systematic uncertainties come from the determination of the beam
intensity, which ranged from 2$\times$10$^3$ to 3$\times$10$^6$ 
$\gamma$/s, with an error bar under 5\%, as discussed 
in \cite{Carson}. The dead 
time (see figure \ref{fig:pressureProfile}), determined to be 2.1 
seconds, had a systematic error in the range from $\pm$2\% up 
to $\pm$15\% for measurements at the highest count rate achieved. 
The length of the liquid target irradiated by the beam was determined 
to be 3.0 $\pm$ 0.1 cm. The uncertainty was mainly determined by the 
position of the $\gamma$-ray beam with respect to the center of the target.
This effect contributed a 3\% systematic error in the determination
of the measured cross sections. The detection efficiency (discussed above) 
contributes a systematic uncertainty that is negligible compared to 
other systematic effects.

\section{Summary}
The long life of stars can be understood in the context of the strong Coulomb barriers which prevent nuclei from fusing. It is this same reason that complicates measurements of very small reaction cross sections in the laboratory. Therefore, new state-of-the-art detectors have to be developed to deal with the extremely low count rates. Since bubble chambers allow us to use liquid target material the density is much higher when compared to gas targets which is advantageous for measuring the small cross sections of time-inverse ($\alpha$, $\gamma$) reactions. The bubble chamber described in this paper is one of the possible approaches. Since these detectors are practically insensitive to $\gamma$-rays only the charged particle reaction products are detected in the active liquid.

We have constructed a device for performing the measurements and established the technique. 
Future facilities for producing $\gamma$-rays using inverse Compton Scattering will benefit from 
results obtained in this work. In particular, upcoming facilities using external lasers that do not have 
significant residual gas in the path of the electron beam and in the same direction of the $\gamma$ ray 
beam will allow measuring the low cross sections relevant to nuclear astrophysics as high energy 
Bremsstrahlung radiation will be significantly reduced. 

This feature is required regardless of the particle identification method or kinematic information obtained 
by a detector trying to measure these small cross sections.

\section{Acknowledgement}
This work was supported by the US Department of Energy, Office of Nuclear Physics, under contract No. DE-AC02-06CH11357. We want to thank the operating group at HI$\gamma$S for providing the high quality beams. Discussions with Prof. Ying K. Wu and Dr. Stepan F. Mikhailov about the source of the background are appreciated. We also thank Sebastian Rehm for writing the LabView computer code used in the acquisition of the experimental data.

\end{document}